\documentclass[ reprint, amsmath,amssymb, aps,]{revtex4-2}

\usepackage{graphicx}
\usepackage{dcolumn}
\usepackage{bm}
\usepackage{comment}
\usepackage{fontenc}
\usepackage{color}
\usepackage{soul}

\usepackage{hyperref}

\begin{document}

\preprint{APS/123-QED}

\title{Nuclear Drip Line and the Composition of Supernova Matter}

\author{\large S. Maity$^{1,2}$}
\email{maity.s@vecc.gov.in}
\author{\large S. Mallik$^{1,2}$}
\email{swagato@vecc.gov.in}
\affiliation{\mbox{$^1$Variable Energy Cyclotron Centre, 1/AF Bidhan Nagar, Kolkata-\textsf{700064}, India}}
\affiliation{\mbox{$^2$Homi Bhabha National Institute, Training School Complex, Anushakti Nagar, Mumbai-\textsf{400094}, India}}

\begin{abstract}
The nuclear drip line plays a crucial role in determining the composition of matter under extreme astrophysical conditions. In core-collapse supernovae and neutron-star crusts, matter is driven far from saturation density and nuclear stability; nuclei coexist with a sea of free neutrons, an effect that is present even at zero temperature in neutron-star crusts and becomes more pronounced in the hotter, neutron-rich supernova environment. This makes a careful treatment of drip-line physics essential for a realistic description of the equation of state and composition. In this work, the influence of the nuclear drip line on the baryonic composition of supernova matter is investigated within the framework of nuclear statistical equilibrium (NSE). The composition is evaluated in terms of free nucleons, light clusters, and heavy nuclei at finite temperature and global sub-saturation densities. The results indicate that, at low proton fractions and higher densities, the inclusion of nuclei beyond the drip line enhances the formation of extremely neutron-rich light clusters, leading to a significant reduction in the free-neutron density and the charge fraction of heavy nuclei. These findings demonstrate that drip-line physics has a significant impact on the composition of supernova matter and should be carefully incorporated in supernova modeling and nucleosynthesis studies.

\end{abstract}

\maketitle

\section{Introduction}

The nuclear drip line defines the limits of nuclear stability and represents one of the central challenges of modern nuclear physics. While the proton drip line has been mapped experimentally over much of the nuclear chart, the neutron drip line remains largely unconstrained due to the combined effects of weak binding, pairing correlations, deformation, and coupling to the particle continuum \cite{Erler2012}. Predictions of the neutron drip line vary substantially among theoretical approaches, including Skyrme energy-density functionals, relativistic mean-field models, and microscopic ab initio calculations, reflecting persistent uncertainties in isovector interactions and many-body correlations \cite{Dobaczewski2007}.\\
\indent
These uncertainties are not only of fundamental interest for nuclear physics but also play a crucial role in nuclear astrophysics, where matter is driven far from stability under extreme thermodynamic conditions. In core-collapse supernovae and proto–neutron stars, matter spans wide ranges of temperature, baryonic density, and proton fraction, enabling the formation of highly neutron-rich nuclei and exotic light clusters which are inaccessible in laboratory experiments \cite{Janka2012,Burrows2013,Furusawa2023,Pinedo}. In neutron-rich astrophysical environments such as supernova envelopes and neutron-star crusts, nuclear clusters coexist with an ambient gas of unbound neutrons. While this coexistence already occurs at zero temperature in neutron-star crusts, thermal effects in supernova matter further enhance the population of weakly bound and neutron-rich nuclei. Under such conditions, the location and treatment of the nuclear drip line become critical for determining the composition and thermodynamic properties of stellar matter. The resulting nuclear composition strongly affects the equation of state (EOS), neutrino opacities, and weak-interaction rates, thereby influencing collapse dynamics, shock revival, and neutrino-driven nucleosynthesis \cite{Langanke2003,MartinezPinedo2012,Hix2003}.\\
\indent
At sub-saturation densities and moderate temperatures, nuclear statistical equilibrium (NSE) provides a natural framework for describing matter as an ensemble of free nucleons, light clusters (atomic number $Z\leq 2$), and heavy nuclei ($Z>2$) in thermal and chemical equilibrium \cite{Bethe1990,Horowitz,Mallik_NuclAstro1}. NSE-based descriptions form the foundation of many modern supernova EOS models \cite{LS,Shen1998,Hempel2010,Typel2010,Raduta2010,Furusawa2011}, which demonstrate that nuclear clustering persists over wide regions of the supernova phase diagram and that light clusters and heavy nuclei coexist with free nucleons well below saturation density.\\
\indent
Light clusters play a significant role in the thermodynamic properties and neutrino interactions of supernova matter \cite{Typel2010,Horowitz}. Quantum statistical approaches and generalized relativistic mean-field models indicate that deuterons, tritons, helions, and alpha particles survive up to densities of order $0.01$–$0.1\rho_0$, modifying the pressure, entropy, and neutrino scattering rates \cite{Ropke2011,Furusawa2017}. Heavy nuclei, on the other hand, dominate electron-capture processes during collapse and strongly influence the deleptonization history of the core \cite{Langanke2003,Furusawa2017b}.\\
\indent
Despite significant progress, a major uncertainty in NSE-based EOS models concerns the treatment of extremely neutron-rich nuclei near and beyond the conventional neutron drip line. Most models rely on experimental mass tables supplemented by theoretical mass models, which necessarily impose practical limits on the nuclear chart \cite{Audi2017}. As a result, weakly bound and continuum-coupled nuclei may be under-represented, particularly in neutron-rich environments relevant for supernova matter. Moreover, a consistent treatment of continuum correlations and the avoidance of double counting between bound and scattering states require explicit subtraction schemes, which are not always implemented in phenomenological NSE formulations \cite{Tubbs,Mallik_NuclAstro1}.\\
\indent
Recent advances in radioactive ion beam (RIB) facilities worldwide have enabled systematic exploration of nuclei far from stability, providing unprecedented access to extremely neutron-rich systems near and beyond the drip lines \cite{Chakrabarty,FRIB,RIBF}. From a nuclear-physics perspective, extending the drip lines increases the available phase space for cluster formation and enhances the role of continuum coupling in stabilizing neutron-rich systems. From an astrophysical viewpoint, this effectively provides additional channels for storing neutrons in bound states, thereby modifying the partitioning of baryon number and isospin between bound clusters and free nucleons. Such changes are expected to affect not only the EOS but also weak-interaction rates, since electron capture and neutrino scattering depend sensitively on nuclear charge distributions and isospin asymmetry \cite{Langanke2003,MartinezPinedo2012}.\\
\indent
In this work, the impact of nuclear drip-line uncertainties on the composition of supernova matter is investigated within a continuum-corrected extended NSE framework \cite{Mallik_NuclAstro1}. The model incorporates self-consistent mean-field interactions for both free nucleons and clusters, finite-size effects through surface and Coulomb contributions, and a subtraction scheme to avoid double counting of continuum states. To quantify the role of the drip line, three nuclear ensembles are considered: one restricted to nuclei within the standard experimental drip line, one allowing moderately extended neutron-rich isotopes, and one including all energetically bound nuclei predicted by the underlying nuclear functional. The sensitivity of the NSE composition to drip-line extension is analyzed as a function of global proton fraction, baryonic density, and temperature, which control, respectively, the neutron richness, clustering strength, and thermal dissolution of nuclei. In addition to conventional mass fractions of free nucleons, light clusters, and heavy nuclei, two bound observables—the bound charge number and the bound isospin asymmetry—are introduced to characterize how charge and neutron excess are stored in clusters and to assess how nuclear-structure uncertainties propagate into quantities relevant for weak-interaction processes and neutrino transport in supernova environments.\\
\indent
The paper is organized as follows. The model is described in Sec.~\ref{sec:NSE}, the results and their discussion are presented in Sec.~\ref{sec:Results}, and the conclusions are given in Sec.~\ref{sec:Conclusions}.
\section{Nuclear statistical equilibrium (NSE) model} \label{sec:NSE}
In the NSE model \cite{Mallik_NuclAstro1,Mallik_NuclAstro2}, the densities of clusters and free nucleons at a given global baryonic  density $\rho_B=\rho_p+\rho_n$, proton fraction $y_p=\rho_p/\rho_B$, and temperature $T$ are obtained within the grand-canonical formalism and are given by,
\begin{eqnarray*}
y_p\rho_{\scriptscriptstyle B} &=& \Big(\frac{2\pi m^*_{p,g}T}{h^2}\Big)^{\frac{3}{2}} F_{1/2}(\eta_{p,g}) + \sum_{Z,N} Z\,\rho_{\scriptscriptstyle Z,N}, \\
(1-y_p)\rho_{\scriptscriptstyle B} &=& \Big(\frac{2\pi m^*_{n,g}T}{h^2}\Big)^{\frac{3}{2}} F_{1/2}(\eta_{n,g}) + \sum_{Z,N} N\,\rho_{\scriptscriptstyle Z,N},
\label{eq:NSE_np}
\end{eqnarray*}
where the first terms on the right-hand side represent the contributions from the free proton ($\rho_{p,g}$) and neutron ($\rho_{n,g}$) gases, respectively, while the second terms account for the number densities of bound clusters ($\rho_{Z,N}$) with proton number $Z$ and neutron number $N$. The quantities $m^*_{p,g}$ and $m^*_{n,g}$ are the effective masses of the free proton and neutron gases, and $\eta_{p,g}=(\mu_p-U_{p,g})/T$ and $\eta_{n,g}=(\mu_n-U_{n,g})/T$ are the corresponding effective chemical potentials within the density-dependent self-consistent mean fields $U_{p,g}$ and $U_{n,g}$, respectively.

The number density of clusters with $Z$ protons and $N$ neutrons ($A=Z+N$) is given by
\begin{eqnarray*}
\rho_{\scriptscriptstyle Z,N}&=&(1-u_c)\Big(\frac{2\pi mT}{h^2}\Big)^{\frac{3}{2}} (Z+N)^{\frac{3}{2}}\nonumber\\
&&\times\phi_{\scriptscriptstyle Z,N}
\exp\!\left[\beta(\mu_p Z+\mu_n N)\right],
\label{eq:cluster_density}
\end{eqnarray*}
where $m$ is the bare nucleon mass, $\phi_{\scriptscriptstyle Z,N}$ is the internal partition function, and $\beta=1/T$. The factor $(1-u_c)$ accounts for excluded volume, reducing the space available for the center-of-mass motion of each cluster, with $u_c = V_c^{\text{tot}}/V^{\text{tot}}$, where $V_c^{\text{tot}}$ and $V^{\text{tot}}$ are the total cluster volume and the total system volume, respectively.

For light clusters (H and He isotopes), no excited states are considered and
$\phi_{\scriptscriptstyle Z,N}=(2J_{\scriptscriptstyle Z,N}+1)\exp(-\beta E_{\scriptscriptstyle Z,N}^{\text{gr}})$,
where $E_{\scriptscriptstyle Z,N}^{\text{gr}}$ and $J_{\scriptscriptstyle Z,N}$ are the energy and spin degeneracy of the ground state respectively. For heavy clusters, the internal partition function is written as
\begin{eqnarray*}
\phi_{\scriptscriptstyle Z,N}=
\exp\!\left[-\frac{F_{\scriptscriptstyle Z,N}^{\text{bulk}}
+F_{\scriptscriptstyle Z,N}^{\text{surf}}
+F_{\scriptscriptstyle Z,N}^{\text{coul}}}{T}\right],
\label{eq:Internal}
\end{eqnarray*}
where $F_{Z,N}^{\text{bulk}}$ is the bulk contribution to the Helmholtz free energy, corresponding to a cluster volume $V_c=A/\rho_c$ of nuclear matter at density
$\rho_c=\rho_{n,c}+\rho_{p,c}$ (with $\rho_{n,c}$ and $\rho_{p,c}$ the neutron and proton densities inside the cluster) and isospin asymmetry
$\delta_c=(\rho_{n,c}-\rho_{p,c})/\rho_c$.

The evaluation of the cluster free energy involves the well-known issue of double counting continuum states in the partition function \cite{Tubbs}. To address this issue, a continuum-subtraction procedure was introduced in Ref.~\cite{Mallik_NuclAstro1}. The continuum-subtracted bulk Helmholtz free energy is
\begin{eqnarray}
F_{Z,N}^{\text{bulk}} &=& V_c \Big[ v(\rho_c,\delta_c)-v(\rho_g,\delta_g) \Big] \nonumber\\
&-& V_c \sum_{q=p,n} \left( U_{q,c}\rho_{q,c}-U_{q,g}\rho_{q,g} \right) \nonumber\\
&-& \frac{2V_c}{3}\Big[ \sum_{q=p,n}(\xi_{q,c}-\xi_{q,g}) + \mu_p Z + \mu_n N \Big],
\end{eqnarray}
where $\eta_{q,c}=(\mu_q-U_{q,c})/T$ are the effective chemical potentials of nucleons inside the cluster within the mean field $U_{q,c}$, and
\[
\xi_{q,c}=\frac{3h^2}{2\pi m^*_{q,c}}
\Big(\frac{2\pi m^*_{q,c}T}{h^2}\Big)^{5/2}
F_{3/2}(\eta_{q,c})
\]
is the kinetic energy density of nucleons of type $q=p,n$ in uniform matter at density $\rho_c$ and asymmetry $\delta_c$.
The quantities $v(\rho_c,\delta_c)$ and $v(\rho_g,\delta_g)$ are the potential energy densities of the bound cluster and of the free nucleon gas
($\rho_g=\rho_{n,g}+\rho_{p,g}$ and $\delta_g=(\rho_{n,g}-\rho_{p,g})/\rho_g$), respectively.
The bulk cluster functional and the corresponding mean fields are obtained from meta-modelling of the EoS \cite{Margueron2018a} using the SLy5 parameters \cite{Chabanat}. Further details are given in Appendix A of Ref.~\cite{Mallik_NuclAstro1}.

Finite-size effects are included through the surface contribution to the Helmholtz free energy,
\begin{eqnarray}
F_{Z,N}^{\text{surf}} &=& 4\pi r_c^2 A_c^{2/3}\, \sigma(y_{p,c},T),
\label{Surface}
\end{eqnarray}
where $A_c=(Z+N)+(\rho_{p,g}+\rho_{n,g})V_c$,
$r_c=\big[3/(4\pi\rho_c)\big]^{1/3}$,
$y_{p,c}=Z/(Z+N)$, and $\sigma(y_{p,c},T)$ is the surface tension at temperature $T$ and proton fraction of the bound cluster $y_{p,c}$ \cite{Mallik_NuclAstro1,LS,Carreau2019}.
The Coulomb free energy (assumed temperature independent) is given by
\begin{equation}
F_{Z,N}^{\text{coul}}=
\frac{3}{5}\frac{e^2 Z^2}{4\pi\epsilon_0}
(1-f_{WS})
\left(\frac{4\pi}{3V_c}\right)^{1/3},
\end{equation}
where $f_{WS}$ is the Wigner--Seitz correction factor accounting for long-range Coulomb interactions in the statistical ensemble \cite{Raduta2015}.
\begin{figure}[!h]
\centering
\vspace{-4.5cm}
\hspace*{1.8cm}\includegraphics[trim=0cm 0cm 0cm 1cm, clip, natwidth=1072,natheight=2462, height=0.75\textheight]{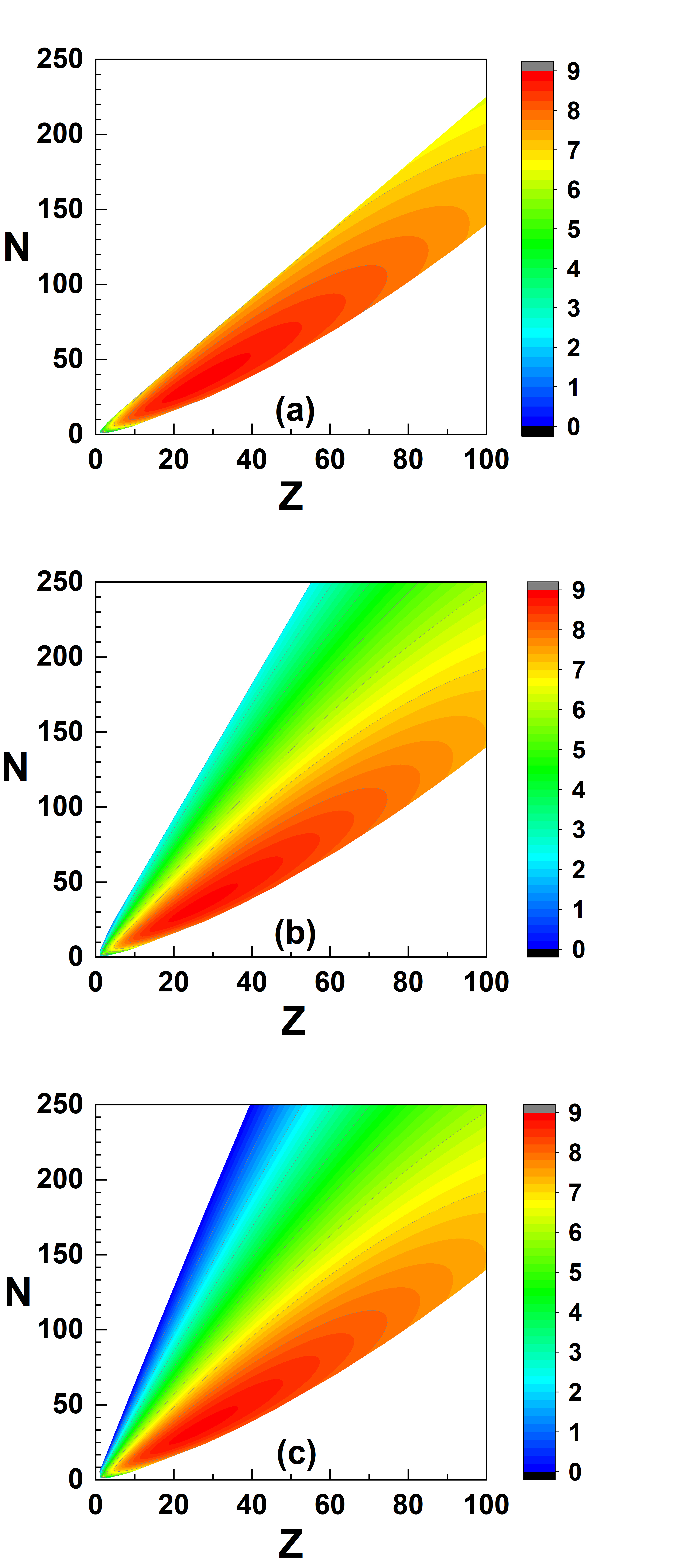}
\caption{Nuclear species included in the three cluster ensembles considered in the present NSE calculations. The distributions are shown in the $Z-N$ plane. The color contour represents the binding energy of the corresponding nuclei. Panels (a), (b), and (c) correspond to the standard drip-line ensemble, the extended drip-line ensemble, and the ensemble containing all energetically bound isotopes, respectively.}
\label{Nuclear_landscapes}
\end{figure}
\section{Results and Discussion}\label{sec:Results}
The present investigation builds upon earlier NSE-based studies of warm stellar and supernova matter, which have demonstrated that the nuclear composition is highly sensitive to thermodynamic conditions and to the treatment of cluster degrees of freedom \cite{LS,Shen1998,Hempel2010,Typel2010,Furusawa2011}. In particular, these works have highlighted the important role of light clusters and intermediate-mass nuclei in shaping the neutron–proton balance and weak-interaction rates during core-collapse evolution \cite{Raduta2015,Arcones2008}. Extending this line of research, the present study focuses specifically on the influence of nuclear drip-line physics on the composition of supernova matter. To this end, continuum-corrected NSE calculations are performed for three distinct nuclear ensembles: (i) $^{2,3}$H, $^{3,4,5,6}$He, are included, and for $Z>2$ only isotopes within the standard drip line are considered (determined from nuclear binding within the compressible liquid-drop approximation using the SLy5 EoS, as described in the Appendix of Ref. \cite{Mallik26}); (ii) an extended drip line is considered, allowing heavy clusters ($Z>2$) with neutron numbers up to twice those of standard drip line, along with H and He isotopes with mass numbers up to 5 and 8, respectively; and (iii) all energetically bound isotopes are included. The bound clusters considered in the three cases are displayed in Fig. 1, where the color contour represents the binding energies of the corresponding nuclei. Although the nuclear landscape is conventionally represented in the $N-Z$ plane, here it is shown in the $Z-N$ plane because the three ensembles mainly differ in the extent of neutron richness allowed for a given atomic number. Although, in principle, the same prescription can also be employed to extend the proton drip line simultaneously, such an extension has a negligible influence on the composition of the extremely neutron-rich supernova matter considered in the present work, while substantially increasing the computational cost. In the following, the impact of nuclear drip line physics on the composition of supernova matter is systematically explored by varying the global proton fraction, baryonic density, and temperature. These three thermodynamic parameters govern the degree of neutron richness, clustering, and thermal dissolution, respectively, and together provide a comprehensive characterization of supernova matter.

\subsection{Global proton-fraction dependence}
\begin{figure}[!b]
\begin{center}
\includegraphics[width=0.7\columnwidth]{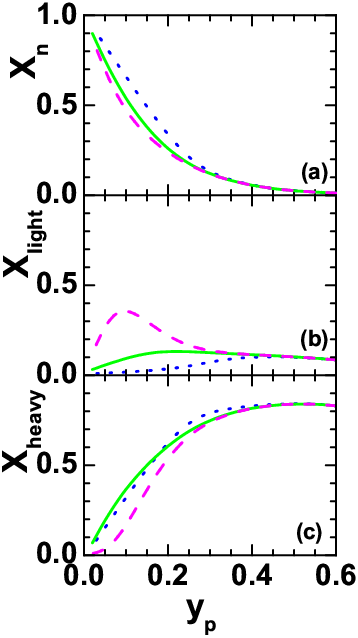}
\caption{Global proton-fraction dependence of the mass fractions of (a) free neutrons, (b) light clusters, and (c) heavy clusters, obtained from NSE calculations at $T=5$ MeV and $\rho_B/\rho_0=0.1$. The results are shown for (i) nuclei within the standard drip line (blue dotted lines), (ii) an extended drip line for heavy clusters with neutron numbers up to twice those of standard nuclei, and H and He isotopes with mass numbers up to 5 and 8, respectively (green solid lines), and (iii) all possible nuclei (magenta dashed lines).}
\label{Mass_fraction_Proton_fraction_Dependence}
\end{center}
\end{figure}
The dependence of the mass fractions on the global proton fraction is studied at a constant temperature of $T=5$ MeV and a global baryon density of $\rho_B=0.1\rho_0$, as shown in Fig.~\ref{Mass_fraction_Proton_fraction_Dependence}. Conventionally, the mass fractions of free neutrons ($X_n$), light clusters ($X_{\text{light}}$), and heavy clusters ($X_{\text{heavy}}$) are defined as\\
\begin{eqnarray}
X_{n} &=& \frac{\rho_{0,1}}{\sum\limits_{Z}\sum\limits_{N}(Z+N)\rho_{Z,N}}, \\
X_{\text{light}} &=& \frac{\sum\limits_{Z=1,2}\sum\limits_{N>0}(Z+N)\rho_{Z,N}}{\sum\limits_{Z}\sum\limits_{N}(Z+N)\rho_{Z,N}}, \\
X_{\text{heavy}} &=& \frac{\sum\limits_{Z>2}\sum\limits_{N}(Z+N)\rho_{Z,N}}{\sum\limits_{Z}\sum\limits_{N}(Z+N)\rho_{Z,N}} .
\end{eqnarray}
Here, the sum in the denominator runs over free nucleons and all possible clusters, while for a given atomic number $Z$, all allowed isotopes with different neutron numbers are included. When nuclei beyond the standard drip line are considered, the free-neutron mass fraction is reduced, particularly at low proton fractions, due to the enhanced formation of extremely neutron-rich clusters. The inclusion of very neutron-rich hydrogen and helium isotopes plays a particularly important role at low proton fractions, reducing not only the free-neutron component but also the contribution from heavy clusters. A comparable reduction of the free-neutron component with decreasing proton fraction has been reported in NSE and generalized relativistic mean field studies of supernova matter, where neutron-rich clusters act as effective neutron reservoirs \cite{Hempel2010,Typel2010,Furusawa2017,Hempel2012}. The present results show that this mechanism is significantly strengthened when nuclei beyond the standard drip line are included, underscoring the importance of an extended nuclear ensemble under extremely neutron-rich conditions.\\
\begin{figure}[!h]
\begin{center}
\includegraphics[width=\columnwidth]{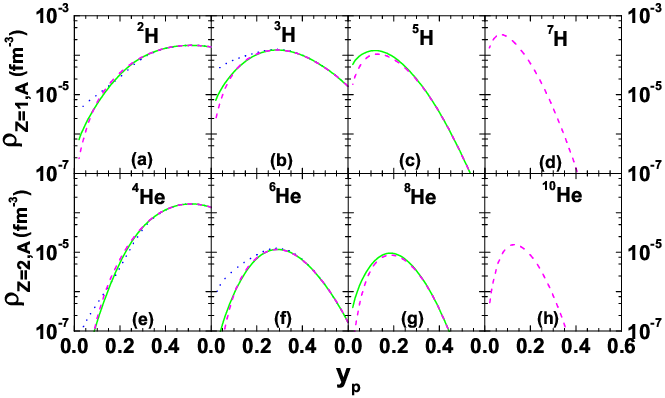}
\caption{Global proton-fraction dependence of densities of selected H (upper panels) and He (lower panels) isotopes calculated within the NSE framework at $T=5$ MeV and global proton fraction $0.2$ at $T=5$ MeV and $\rho_B/\rho_0=0.1$. The results are shown for (i) nuclei within the standard drip line (blue dotted lines), (ii) an extended drip line for heavy clusters with neutron numbers up to twice those of standard nuclei, and H and He isotopes with mass numbers up to 5 and 8, respectively (green solid lines), and (iii) all possible nuclei (magenta dashed lines).}
\label{H_He_Clusters_Proton_Fraction_Dependence}
\end{center}
\end{figure}
\indent
To elucidate this behavior, Fig.~\ref{H_He_Clusters_Proton_Fraction_Dependence} shows the global proton-fraction dependence of the densities of selected H and He isotopes under the same thermodynamic conditions. Exotic light nuclei such as $^{6,7}$H and $^{8,10}$He become abundant in neutron-rich supernova matter at low proton fractions, although they remain extremely difficult to access in heavy-ion experiments. The gradual population of these neutron-rich light clusters suppresses the densities of comparatively less neutron-rich isotopes of the same elements.\\
\begin{figure}[!b]
\begin{center}
\includegraphics[width=\columnwidth]{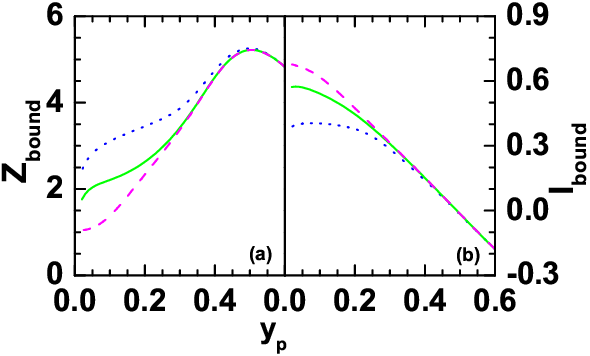}
\caption{Global proton-fraction dependence of the mass fractions of (a) bound charge number ($Z_{bound}$)  and (b) bound isospin asymmetry ($I_{bound}$), obtained from NSE calculations at $T=5$ MeV and $\rho_B/\rho_0=0.1$. Results are shown for nuclei within the standard drip line (blue dotted lines), an extended drip line including neutron-rich heavy clusters and H and He isotopes up to mass numbers 5 and 8, respectively (green solid lines), and all possible nuclei (magenta dashed lines).}
\label{Bound_Proton_Fraction_Dependence}
\end{center}
\end{figure}
\indent
Motivated by these observations, the effect of the nuclear drip line on two theoretically important observables for supernova matter is examined: (a) the bound charge number $Z_{\text{bound}}$, which characterizes the effective charge stored in nuclear clusters, and (b) the bound isospin asymmetry $I_{\text{bound}}$, which quantifies the neutron--proton imbalance of bound matter. These observables are defined as
\begin{eqnarray}
Z_{\text{bound}} &=&
\frac{\sum\limits_{Z>0}\sum\limits_{N>0} Z\,\rho_{Z,N}}
{\sum\limits_{Z>0}\sum\limits_{N>0}\rho_{Z,N}}, \\
I_{\text{bound}} &=&
\frac{\sum\limits_{Z>0}\sum\limits_{N>0} \frac{(N-Z)}{(N+Z)}\,\rho_{Z,N}}
{\sum\limits_{Z>0}\sum\limits_{N>0}\rho_{Z,N}} .
\end{eqnarray}
The dependence of $Z_{\text{bound}}$ and $I_{\text{bound}}$ on the global proton fraction is shown in Fig.~\ref{Bound_Proton_Fraction_Dependence}. At moderate proton fractions ($0.4 \le Y_p \le 0.6$), the abundance of extremely neutron-rich isotopes is negligible, and the effect of drip-line extension on both observables remains small. As the proton fraction decreases, neutron-rich light clusters become increasingly abundant, leading to a reduction of $Z_{\text{bound}}$ and an enhancement of $I_{\text{bound}}$. Since these observables are closely linked to electron-capture rates and neutrino emission, they are expected to influence weak-interaction processes in core-collapse supernova environments.\\
\begin{figure}[!h]
\begin{center}
\includegraphics[width=\columnwidth]{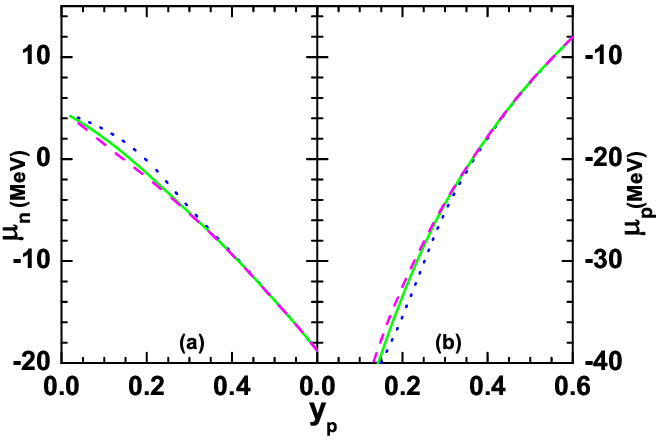}
\caption{Global proton-fraction dependence of the (a) neutron ($\mu_n$) and (b) proton ($\mu_p$) chemical potentials obtained from NSE calculations at $T=5$ MeV and $\rho_B/\rho_0=0.1$. Results are shown for the standard drip-line nuclei (blue dotted lines), the extended drip-line nuclei (green solid lines), and all energetically bound nuclei (magenta dashed lines).}
\label{Chemical_Potential_Proton_Fraction_Dependence}
\end{center}
\end{figure}
\indent
In addition to the cluster abundances and bound observables, the dependence of the neutron and proton chemical potentials on the global proton fraction is also investigated for the three different nuclear ensembles, as shown in Fig.~\ref{Chemical_Potential_Proton_Fraction_Dependence}. For $y_p\geq0.3$, the differences among the three calculations remain relatively small because the contribution of extremely neutron-rich nuclei is limited. For $0.05\leq y_p<0.3$, the inclusion of nuclei beyond the standard drip line allows excess free neutrons to be incorporated into bound states. This reduces the free-neutron component and consequently lowers the neutron chemical potential. The proton chemical potential is also modified indirectly through chemical equilibrium conditions. For $y_p<0.05$, the production of bound clusters becomes strongly suppressed irrespective of the choice of drip-line prescription due to the extremely neutron-rich environment. As a result, the matter composition becomes increasingly dominated by free neutrons, leading to a gradual convergence of the neutron chemical potentials obtained in the different calculations.\\
\indent
While variations in the global proton fraction primarily control the neutron richness of the system and the accessibility of exotic nuclei, the baryonic density determines the strength of nucleon–nucleon correlations and the onset of cluster formation. It is therefore instructive to examine how the nuclear composition and bound observables evolve with increasing density under fixed isospin conditions.
\subsection{Global baryonic density dependence}
\begin{figure}[!h]
\begin{center}
\includegraphics[width=0.7\columnwidth]{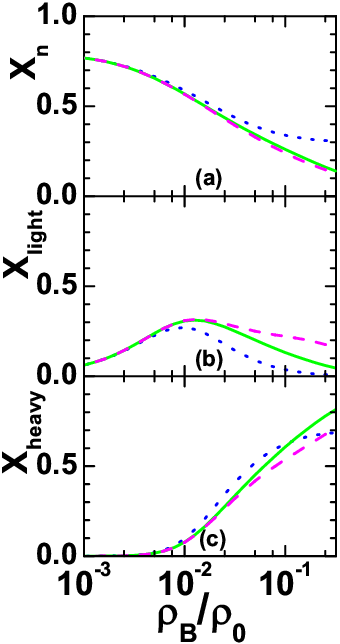}
\caption{Global baryonic density dependence of the mass fractions of (a) free neutrons, (b) light clusters, and (c) heavy clusters, obtained from NSE calculations at $T=5$ MeV and global proton fraction $0.2$. The results are shown for (i) nuclei within the standard drip line (blue dotted lines), (ii) an extended drip line for heavy clusters with neutron numbers up to twice those of standard nuclei, and H and He isotopes with mass numbers up to 5 and 8, respectively (green solid lines), and (iii) all possible nuclei (magenta dashed lines).}
\label{Mass_Fraction_Baryonic_Density_Dependence}
\end{center}
\end{figure}
The dependence of the NSE composition on the global baryon density is studied at a fixed temperature of $T=5$ MeV and a global proton fraction of $Y_p=0.2$. Figure~\ref{Mass_Fraction_Baryonic_Density_Dependence} shows the density evolution of the mass fractions of free neutrons, light clusters, and heavy clusters over the sub-saturation density range relevant to supernova matter. At low baryonic densities, the composition is dominated by free neutrons, reflecting the neutron-rich environment. With increasing density, nucleon--nucleon correlations become stronger, favoring cluster formation and resulting in a systematic reduction of the free-neutron mass fraction accompanied by enhanced production of both light and heavy clusters.\\
\indent
The sensitivity of this evolution to the treatment of the nuclear drip line is evident. The inclusion of neutron-rich nuclei beyond the standard drip line allows extremely neutron-rich clusters to become energetically accessible at higher densities, leading to a stronger suppression of the free-neutron component. The enhanced formation of neutron-rich light clusters also reduces the relative contribution of heavy clusters, indicating a redistribution of baryon number among different cluster species.\\
\begin{figure}[!h]
\begin{center}
\includegraphics[width=\columnwidth]{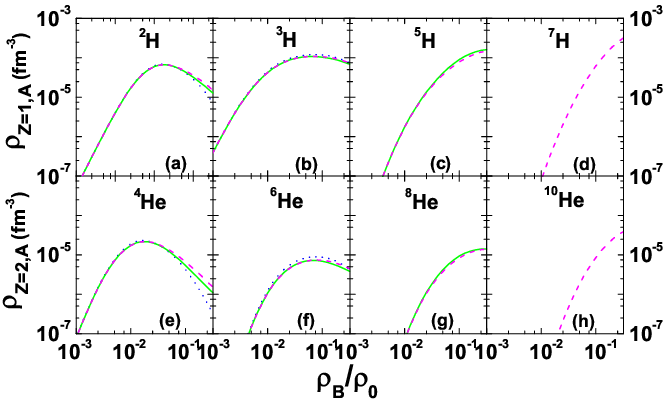}
\caption{Global baryonic density dependence of selected H (upper panels) and He (lower panels) isotopes calculated within the NSE framework at $T=5$ MeV and global proton fraction $0.2$. The results are shown for (i) nuclei within the standard drip line (blue dotted lines), (ii) an extended drip line for heavy clusters with neutron numbers up to twice those of standard nuclei, and H and He isotopes with mass numbers up to 5 and 8, respectively (green solid lines), and (iii) all possible nuclei (magenta dashed lines).}
\label{H_He_Clusters_Baryonic_Density_Dependence}
\end{center}
\end{figure}
\indent
The density dependence of selected hydrogen and helium isotopes is shown in Fig.~\ref{H_He_Clusters_Baryonic_Density_Dependence}. In the standard drip-line description, only weakly neutron-rich isotopes such as $^{2,3}$H and $^{4,5,6}$He are populated, with densities increasing moderately with baryon density. When neutron-rich nuclei are included, exotic isotopes such as $^{7}$H and $^{8}$He appear at higher densities, and in the most inclusive case extremely neutron-rich species such as $^{10}$He acquire sizable densities. These isotopes suppress less neutron-rich species and further reduce the free-neutron content.\\
\begin{figure}[!h]
\begin{center}
\includegraphics[width=\columnwidth]{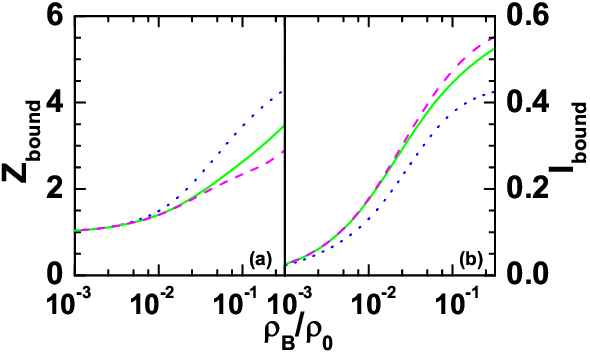}
\caption{Global baryonic density dependence of the mass fractions of (a) bound charge number ($Z_{bound}$)  and (b) bound isospin asymmetry ($I_{bound}$), obtained from NSE calculations at $T=5$ MeV and global proton fraction $0.2$. Results are shown for nuclei within the standard drip line (blue dotted lines), an extended drip line including neutron-rich heavy clusters and H and He isotopes up to mass numbers 5 and 8, respectively (green solid lines), and all possible nuclei (magenta dashed lines).}
\label{Bound_Baryonic_Density_Dependence}
\end{center}
\end{figure}
\indent
The corresponding density dependence of the bound observables is shown in Fig.~\ref{Bound_Baryonic_Density_Dependence}. At low densities, where clustering is weak, both $Z_{\text{bound}}$ and $I_{\text{bound}}$ exhibit only a mild dependence on the nuclear ensemble. With increasing density, $Z_{\text{bound}}$ decreases and $I_{\text{bound}}$ increases in extended drip-line calculations, reflecting the growing dominance of neutron-rich clusters and the enhanced neutron excess stored in bound nuclei. The enhanced neutron excess stored in bound clusters at higher densities is consistent with previous studies showing that cluster asymmetry increases toward the inner crust of neutron stars and in late-stage supernova matter \cite{Furusawa2017,Newton2013}.\\
\begin{figure}[!b]
\begin{center}
\includegraphics[width=\columnwidth]{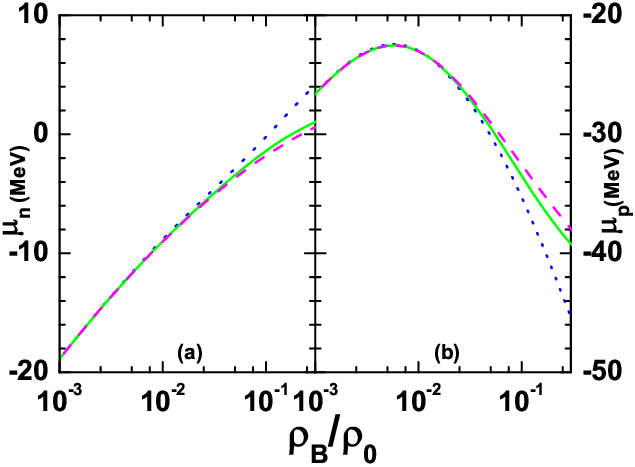}
\caption{Global baryonic density dependence of the (a) neutron ($\mu_n$) and (b) proton ($\mu_p$) chemical potentials obtained from NSE calculations at $T=5$ MeV and global proton fraction $0.2$. Results are shown for the standard drip-line nuclei (blue dotted lines), the extended drip-line nuclei (green solid lines), and all energetically bound nuclei (magenta dashed lines).}
\label{Chemical_Potential_Density_Dependence}
\end{center}
\end{figure}
\indent
Fig.~\ref{Chemical_Potential_Density_Dependence} shows the baryonic density dependence of the neutron and proton chemical potentials for the three nuclear ensembles. At low densities, where the matter composition is dominated by free nucleons, the differences among the calculations remain relatively small. With increasing density, cluster correlations become stronger and neutron-rich nuclei acquire significant abundances, particularly in the extended drip-line calculations. Consequently, the neutron chemical potential decreases more rapidly with density when nuclei beyond the standard drip line are included, reflecting the enhanced binding of excess neutrons into highly asymmetric clusters. The proton chemical potential also exhibits a baryonic density dependence on the nuclear ensemble due to the modified chemical equilibrium among free nucleons and clusters. The sensitivity of the chemical potentials to the drip-line prescription at higher densities suggests that extremely neutron-rich nuclei may influence neutrino opacities and thermal transport properties in dense supernova matter and proto-neutron-star environments.\\
\indent
At a given density and proton fraction, thermal effects further regulate the balance between bound clusters and free nucleons by controlling nuclear stability and entropy. The temperature dependence of the NSE composition thus provides crucial insight into the competition between binding and thermal dissolution in supernova matter.
\subsection{Temperature dependence}
\begin{figure}[!h]
\begin{center}
\includegraphics[width=0.7\columnwidth]{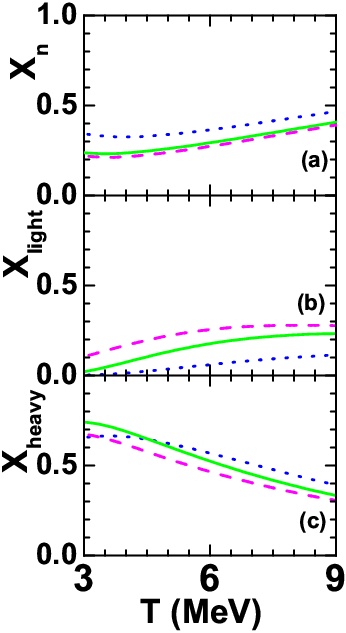}
\caption{Temperature dependence of the mass fractions of (a) free neutrons, (b) light clusters, and (c) heavy clusters, obtained from NSE calculations at $\rho_B/\rho_0=0.1$ and $y_p=0.2$. The results are shown for (i) nuclei within the standard drip line (blue dotted lines), (ii) an extended drip line for heavy clusters with neutron numbers up to twice those of standard nuclei, and H and He isotopes with mass numbers up to 5 and 8, respectively (green solid lines), and (iii) all possible nuclei (magenta dashed lines).}
\label{Mass_Fraction_Temperature_Dependence}
\end{center}
\end{figure}
Fig.~\ref{Mass_Fraction_Temperature_Dependence} shows the temperature dependence of the NSE composition at $\rho_B/\rho_0=0.1$ and $Y_p=0.2$. At high temperatures, thermal excitation dominates in all cases, leading to the dissolution of nuclear clusters and convergence toward a nucleon-dominated regime with a large free-neutron mass fraction. As the temperature decreases, cluster formation becomes increasingly favorable. In the standard drip-line description, heavy neutron-rich clusters gradually build up, accompanied by a reduction of the free-neutron fraction. The inclusion of neutron-rich nuclei enhances cluster stability over a broader temperature range, leading to systematically lower free-neutron mass fractions and higher light-cluster abundances, with the strongest effect observed when all energetically bound nuclei are included. The dissolution of clusters with increasing temperature is a generic feature of warm nuclear matter in NSE and virial approaches \cite{Horowitz,Hempel2010}. Extending the drip line delays this dissolution by stabilizing neutron-rich light clusters.\\
\indent
At temperatures below $T \simeq 3.5$ MeV, the free-neutron mass fraction exhibits a slight increase with decreasing temperature. This behavior arises from the suppression of entropy-favored light clusters at low temperatures, which releases neutrons that cannot be fully absorbed by increasingly neutron-rich bound nuclei, resulting in a small residual population of free neutrons despite the dominance of heavy clusters.\\
\begin{figure}[!h]
\begin{center}
\includegraphics[width=\columnwidth]{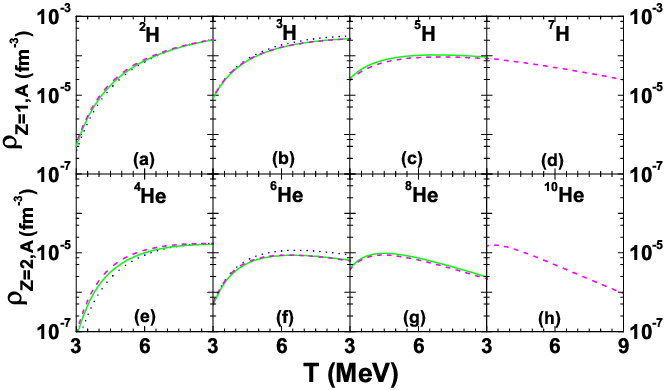}
\caption{Temperature dependence of selected H (upper panels) and He (lower panels) isotopes calculated within the NSE framework at $\rho_B/\rho_0=0.1$ and $y_p=0.2$. The results are shown for (i) nuclei within the standard drip line (blue dotted lines), (ii) an extended drip line for heavy clusters with neutron numbers up to twice those of standard nuclei, and H and He isotopes with mass numbers up to 5 and 8, respectively (green solid lines), and (iii) all possible nuclei (magenta dashed lines).}
\label{H_He_Clusters_Temperature_Dependence}
\end{center}
\end{figure}
\indent
Figure~\ref{H_He_Clusters_Temperature_Dependence} further elucidates the microscopic origin of these trends. The inclusion of neutron-rich nuclei leads to the population of exotic isotopes such as $^{5,7}$H and $^{8,10}$He at low temperatures. These neutron-rich light clusters suppress less neutron-rich species by redistributing neutrons into more asymmetric bound states, highlighting the strong sensitivity of light-cluster abundances to drip-line physics in cold supernova matter.\\
\begin{figure}[!b]
\begin{center}
\includegraphics[width=\columnwidth]{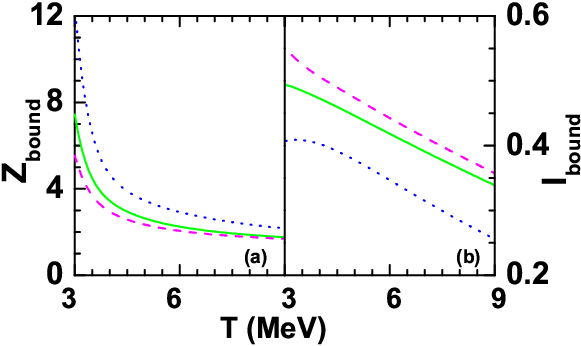}
\caption{Temperature dependence of the mass fractions of (a) bound charge number ($Z_{bound}$)  and (b) bound isospin asymmetry ($I_{bound}$), obtained from NSE calculations at $\rho_B/\rho_0=0.1$ and $y_p=0.2$. Results are shown for nuclei within the standard drip line (blue dotted lines), an extended drip line including neutron-rich heavy clusters and H and He isotopes up to mass numbers 5 and 8, respectively (green solid lines), and all possible nuclei (magenta dashed lines).}
\label{Bound_Temperature_Dependence}
\end{center}
\end{figure}
\indent
Fig.~\ref{Bound_Temperature_Dependence} shows the temperature dependence of $Z_{\text{bound}}$ and $I_{\text{bound}}$. At high temperatures, both observables remain small and exhibit only a weak dependence on the drip-line prescription due to cluster dissolution. As the temperature decreases, $Z_{\text{bound}}$ decreases while $I_{\text{bound}}$ increases, reflecting the enhanced formation of neutron-rich clusters. These effects are most pronounced at low temperatures, emphasizing the crucial role of nuclear structure in determining the composition and weak-interaction properties of cold supernova matter.\\
\begin{figure}[!t]
\begin{center}
\includegraphics[width=0.7\columnwidth]{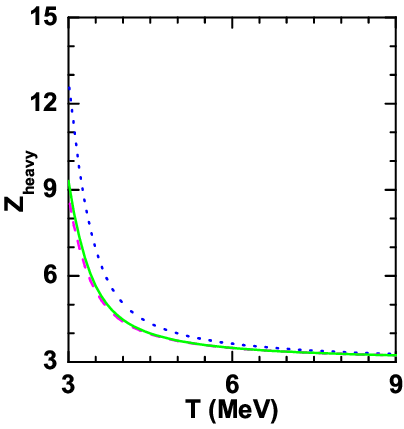}
\caption{Temperature dependence of the average charge of heavy nuclei $Z_{heavy}$, obtained from NSE calculations at $\rho_B/\rho_0=0.1$ and $y_p=0.2$. Results are shown for the standard drip-line nuclei (blue dotted lines), the extended drip-line nuclei (green solid lines), and all energetically bound nuclei (magenta dashed lines).}
\label{Zheavy_Temperature_Dependence}
\end{center}
\end{figure}
\indent
The temperature dependence of the average charge of heavy nuclei $Z_{heavy}$, is also investigated for the three different nuclear ensembles and shown in Fig.~\ref{Zheavy_Temperature_Dependence}. At high temperatures, the differences among the calculations remain small due to thermal dissolution of heavy clusters. As the temperature decreases, $Z_{heavy}$ increases in all three cases because of the enhanced formation of bound nuclei. However, at a given temperature, the significant production of neutron-rich H and He isotopes in the extended drip-line and all-bound-nuclei calculations reduces the relative abundance of heavy clusters, leading to comparatively smaller values of $Z_{heavy}$ than in the standard drip-line case. Since coherent neutrino--nucleus scattering depends sensitively on the properties of heavy nuclei and plays an important role in late-stage proto-neutron-star cooling \cite{Nakazato2018}, these results may have implications for neutrino cooling in proto-neutron stars.\\
\begin{figure}[!b]
\begin{center}
\includegraphics[width=\columnwidth]{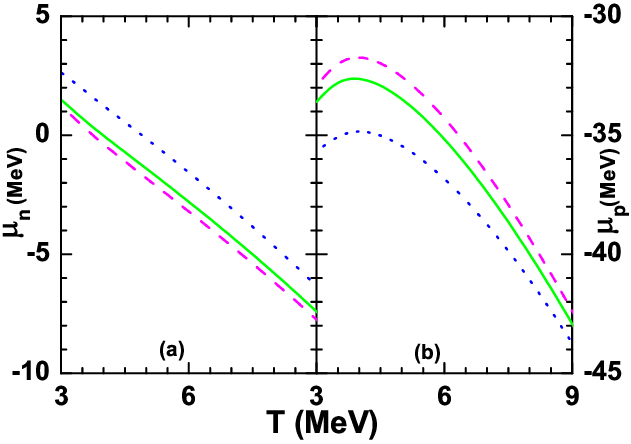}
\caption{Temperature dependence of the of the (a) neutron ($\mu_n$) and (b) proton ($\mu_p$) chemical potentials obtained from NSE calculations at $\rho_B/\rho_0=0.1$ and $y_p=0.2$. Results are shown for nuclei within the standard drip line (blue dotted lines), an extended drip line including neutron-rich heavy clusters and H and He isotopes up to mass numbers 5 and 8, respectively (green solid lines), and all possible nuclei (magenta dashed lines).}
\label{Chemical_Potential_Temperature_Dependence}
\end{center}
\end{figure}
\indent
The temperature dependence of the neutron and proton chemical potentials are shown in Fig.~\ref{Chemical_Potential_Temperature_Dependence}. At high temperatures, thermal excitation suppresses cluster formation and the chemical potentials obtained in the three calculations become nearly identical. As the temperature decreases, cluster formation becomes increasingly favorable, particularly for neutron-rich nuclei beyond the standard drip line. The enhanced population of extremely neutron-rich clusters lowers the neutron chemical potential by transferring excess neutrons from the free nucleon gas into bound states. The differences among the three calculations become most pronounced at low temperatures, where clustering effects are strongest. The proton chemical potential exhibits an opposite trend due to the coupled chemical-equilibrium conditions among nucleons and clusters. These results demonstrate that the treatment of nuclear drip-line physics can significantly influence the thermodynamic properties of cold and moderately warm supernova matter, with possible consequences for neutrino emission and proto-neutron-star cooling.\\
\indent
Finally, these results demonstrate that a proper treatment of nuclear drip-line physics can substantially modify the composition of warm and cold supernova matter over a broad range of densities, temperatures, and proton fractions. Such changes in composition are expected to affect neutrino transport, electron-capture rates, and the thermodynamic evolution of core-collapse supernovae and neutron-star crusts, highlighting the importance of consistently incorporating extremely neutron-rich nuclei in astrophysical equations of state \cite{Furusawa2017b,Typel2014}.\\
\indent
It should also be noted that the largest differences among the three nuclear ensembles are obtained mainly in the high-density region. This region lies inside the neutrino sphere, where neutrinos remain trapped and matter stays close to weak equilibrium. Consequently, the differences among the three nuclear ensembles are expected to have only a modest effect on the overall dynamical evolution of core-collapse supernovae and on the early cooling stage of proto-neutron stars. Nevertheless, the observed modifications of the nuclear composition, heavy-cluster properties, and neutron/proton chemical potentials may still influence local weak-interaction rates, neutrino opacities, and thermal transport properties in dense neutron-rich matter. A quantitative assessment of these effects on proto-neutron-star cooling and supernova evolution would require self-consistent neutrino-radiation hydrodynamics simulations employing extended nuclear ensembles beyond the standard drip line.\\
\indent
In the present study, the in-medium modification of the nuclear binding energy is not considered, which could possibly reduce the statistical weight of loosely bound clusters. The in-medium effect is not significant for heavy nuclei, but is important for the light clusters. Quantum statistical approaches \cite{Roepke2009,Roepke2015,Roepke2021} and recent investigations of chemical equilibrium constants \cite{Mallik_NuclAstro2,QinPRL108,Pais2020} from intermediate energy heavy-ion reactions provide useful information on binding energy shifts of light clusters. In future work, it will be interesting to introduce the in-medium modification of nuclear binding energy in NSE model and to study its effect on supernova matter.
\section{Summary and Conclusions}\label{sec:Conclusions}
The influence of nuclear drip-line physics on the composition of supernova matter has been studied within a continuum-corrected nuclear statistical equilibrium framework. By systematically extending the nuclear ensemble beyond the standard drip line, the sensitivity of matter composition to extreme neutron-rich nuclei is examined over a wide range of proton fractions, baryonic densities, and temperatures relevant to core-collapse environments.\\
\indent
The results demonstrate extending the nuclear ensemble beyond the conventional drip line leads to substantial modifications of the NSE composition under neutron-rich conditions. At low proton fractions and sub-saturation densities, the inclusion of extremely neutron-rich nuclei strongly enhances the formation of exotic light clusters, particularly neutron-rich hydrogen and helium isotopes. These clusters act as efficient neutron reservoirs, resulting in a pronounced reduction of the free-neutron component and a redistribution of baryon number from heavy nuclei toward light clusters. Consequently, the effective charge stored in bound nuclei decreases, while the neutron excess of bound matter increases. The impact of drip-line physics becomes increasingly significant with rising density, where enhanced nucleon–nucleon correlations favor clusterization, and with decreasing temperature, where thermal dissolution is suppressed. In these regimes, extended drip-line treatments delay the dissolution of clusters and stabilize neutron-rich bound states over a broader thermodynamic domain. The resulting increase in the bound isospin asymmetry and the reduction of the bound charge number highlight the sensitivity of composition-related observables to nuclear-structure uncertainties far from stability.\\
\indent
These findings have direct implications for supernova modeling and neutron-star physics. In addition to the composition and bound observables, the neutron and proton chemical potentials also exhibit sensitivity to the treatment of the nuclear drip line, particularly in dense, cold, and neutron-rich matter. The resulting modifications of these thermodynamic quantities may influence neutrino transport, weak-interaction rates, and proto-neutron-star cooling. Since electron-capture rates, neutrino opacities, and weak-interaction processes depend sensitively on nuclear charge distributions and isospin asymmetry, uncertainties in the treatment of the neutron drip line may propagate into the dynamical and thermal evolution of core-collapse supernovae and the structure of neutron-star crusts. The present results therefore underscore the importance of consistently incorporating extremely neutron-rich nuclei and continuum effects in astrophysical equations of state.\\
\indent
Looking ahead, improved experimental constraints on neutron-rich nuclei from next-generation radioactive-beam facilities, together with advances in microscopic nuclear theory and continuum treatments, will be essential for reducing these uncertainties. The present study provides a systematic framework for assessing the astrophysical impact of drip-line physics and highlights the need for extended and self-consistent nuclear ensembles in realistic descriptions of dense, neutron-rich matter.\\
\section{Acknowledgement}
The authors gratefully acknowledge Francesca Gulminelli of LPC Caen for valuable discussions and suggestions.

\end{document}